\documentclass[twocolumn,aps,showpacs,preprintnumbers,nofootinbib]{revtex4}
\usepackage{axodraw}
\usepackage{graphicx}
\usepackage{amsmath}
\usepackage{amssymb}

\newcommand{\mw}[0]{M_{\text{W}}}
\newcommand{\lh}[0]{\Lambda_{\text{H}}}
\newcommand{\gf}[0]{G_{\text{F}}}

\newcommand{\Eq}[1]{Eq.~(\ref{#1})}

\newcommand{\Fig}[1]{Fig.~\ref{#1}}

\begin{document}

%%{\noindent
%%\hfill
%%\begin{flushright}
%%CALT-68-2372\\
%%hep-ph/0309115\\
%%\end{flushright}
%%}

\author{Gary Pr\'{e}zeau, Andriy Kurylov, Marc Kamionkowski, Petr Vogel}
\affiliation{
Division of Physics, Mathematics, and Astronomy, California
Institute of Technology, Pasadena, CA 91125}

\title{New contribution to WIMP-nucleus scattering}

%\date

\begin{abstract}

A weakly-interacting massive particle (WIMP) is perhaps the most
promising candidate for the dark matter in the Galactic halo.
The WIMP detection rate in laboratory searches is fixed by the
cross section for elastic WIMP-nucleus scattering.  Here we
calculate the contribution to this cross section from
two-nucleon currents from pion exchange in the nucleus, and show
that it may in some cases be comparable to the one-nucleon
current that has been considered in prior work, and perhaps help
resolve the discrepancies between the various direct dark-matter
search experiments.  We provide
simple expressions that allow these new contributions to be
included in current calculations.

\end{abstract}


%\narrowtext
\maketitle

%\arabic{figure}
\pagenumbering{arabic}

A neutral, stable, weakly-interacting massive particle (WIMP) with 
mass near the electroweak scale is one of the most 
natural dark-matter candidates \cite{kam-rev,bergstrom} because it has a
cosmological density comparable to that contributed by halo dark
matter. The most widely studied WIMP candidate is the
neutralino---a linear combination of the supersymmetric partners of
the photon, $Z$ and Higgs bosons---in the minimal
supersymmetric extension of the standard model (MSSM). There are, however,
other possibilities, including the sneutrino (the neutrino's
superpartner), heavy neutrinos, Kaluza-Klein modes in models with
universal extra dimensions, {\it etc}.

If such WIMPs exist, one way to observe them is directly via
detection of the ${\cal O}(30\,{\rm keV})$ recoil energy imparted
to a nucleus in a low-background detector when a WIMP
collides with the nucleus \cite{witten85,wasserman}.
The predicted event rate for these searches depends primarily on
the cross section for elastic WIMP-nucleus scattering.
Typically, a WIMP-nucleon interaction is derived  from the
WIMP-quark interaction that appears in the MSSM (or other
WIMP-theory) Lagrangian.  It has long been known that for
supersymmetric models---and recently shown for more general
pointlike WIMPs \cite{kurylov03}---the WIMP can
couple to the nucleon with either a spin-independent (SI) or
spin-dependent (SD) interaction.  Nuclear physics then allows
the WIMP-nucleus cross section to be derived from the SI and SD
WIMP-nucleon interactions.

In this work, we use an effective field-theory description of
nuclear interactions to consider the contribution to
WIMP-nucleus scattering from two-nucleon currents,
heuristically, the coupling of WIMPs to the virtual pions that
hold the nucleus together.  Although this contribution is, as
one might expect, small in many cases, for SI interactions it is
usually non-negligible and sometimes comparable to or even
bigger than the one-nucleon current usually considered.

We begin with the underlying
WIMP-quark interaction Lagrangian, construct effective hadronic
operators, and organize them according to the order in
$q/\lh$ with $\lh\sim 1$~GeV a hadronic scale like the nucleon mass
or the chiral symmetry breaking scale and $q\sim m_\pi$, the pion mass
with $q$ being the momentum transfer or the Fermi momentum of the
nucleons inside the nucleus. 
We then focus on the neutralino to perform an explicit evaluation of
the two-nucleon scattering amplitude and we estimate the range for its
magnitude.

The two-nucleon contributions will have a long range piece through the
exchange of pions 
and a contact piece as shown in the diagrams of \Fig{twobody} where
the WIMP is denoted by ${\cal{X}}$.  The
long range contributions are enhanced compared to the
two-nucleon contact interaction as can be seen from power counting in the
small momentum $q$. To show this, we first organize the ${\cal{X}}$-hadron
vertices in powers of $q$ using chiral perturbation theory
\cite{Gasser:1983yg} and the
fact that the momentum transfered by the WIMP is $\sim 100$~MeV or less for
WIMP masses $\sim 100$~GeV or less.  The
leading order (LO) quark operators should therefore induce effective
WIMP-hadron vertices that do not involve derivatives of the pion fields
or pion mass insertions, (which
are always quadratic and therefore do not contribute at LO or
  next-to-leading order (NLO)) the NLO operators would
involve a single derivative of the pion field, the
next-to-next-to-leading order (NNLO) would involve two derivatives or
pion mass insertions and so-on~\cite{Prezeau:2003xn}.  Thus, counting
pion  propagators as ${\cal{O}}(1/q^2)$ and the pion-nucleon vertex
as ${\cal{O}}(q)$, we find that the matrix element of 
$\text{Fig.~\ref{twobody}(a)}\sim {\cal{O}}(K_{\pi\pi} q^{-2}),
~ \text{Fig.~\ref{twobody}(b,c)} \sim {\cal{O}}(K_{NN\pi}q^{-1}),
~ \text{Fig.~\ref{twobody}(d)} \sim {\cal{O}}(K_{N^4}q^0),$
where the $K_i$ denote the order of the ${\cal{X}}$-hadron
vertices. In general, the LO ${\cal{X}}$-hadron vertex in each
diagram will have $K_i\sim{\cal O}(q^0)$, though in certain cases
symmetry considerations 
of the underlying particle-physics model require that the LO vertex
vanish~\cite{Prezeau:2003xn}.  Thus, the long range 
two-nucleon 
contributions of Figs.~\ref{twobody}(a), and \ref{twobody}(b,c) are
enhanced by $1/q^2$ and $1/q$, respectively, relative to the short-range
operator of Fig.~\ref{twobody}(d).  In what follows, we will at most
consider NLO contributions to the scattering amplitude and will
henceforth neglect the contact term.  Therefore, we consider all
terms in $K_{\pi\pi}$ and $K_{NN\pi}$ to ${\cal{O}}(q)$ and
${\cal{O}}(1)$ respectively.
%%%%%%%%%%%%%%%%%%%%%%%%%%%%%%%%%%%%%%%%%%%%%%%%%%%%%%%%
%\suppressfloats[t]
\begin{center}
\begin{figure*}\caption{Two-nucleon diagrams that contribute to
WIMP-nucleus scattering where the WIMP is generally denoted by ${\cal{X}}$.
Graph (a) is of ${\cal{O}}(1/q^2)$, graphs 
(b) and (c) are of ${\cal{O}}(1/q)$ while the contact term of graph
(d) is of ${\cal{O}}(1)$.  The exchange diagrams are not
included.  The filled circles represent the non-standard model
vertices.}\label{twobody} 
\resizebox{16cm}{!}{\includegraphics*[38,620][578,720]{diagrams.ps}} 
\vspace{-1cm}
\end{figure*}
\end{center}
\vspace{-0.8cm}
%%%%%%%%%%%%%%%%%%%%%%%%%%%%%%%%%%%%%%%%%%%%%%%%%%%%%%%%

We  now  construct the  ${\cal{X}}$-hadron
effective Lagrangian from an underlying ${\cal{X}}$-quark effective
Lagrangian where ${\cal{X}}$ is a general neutral WIMP field with
arbitrary spin, isospin, and parity (note that we are not yet
specializing to the neutralino).  The most general set of
CP-conserving interactions between WIMP and quark currents is
%%%%%%%%%%%%%%%%%%%%%%%%%%%%%%%%%%%%%%%%%%%%%%%%%%%%%%%%%%
\begin{eqnarray}\label{quarklag}
{\cal{L}}_{{\cal{X}}q} =
\gf
\sum\limits_{q}
\left[ a_1^q{\cal{S}}\bar{q}q
  +a_2^q{\cal{P}}\bar{q}\gamma_5q  
\right.
~~~~~~~~~~~~~~~~~~~~~~~~~~& &
\nonumber
\\
+{\cal{V}}^\mu  
\bar{q}\gamma_\mu(a_3^q+a_4^q\gamma_5) q 
+{\cal{A}}^\mu 
\bar{q}\gamma_\mu(a_5^q+a_6^q\gamma_5) q 
~~~& &
\nonumber
\\
\left.
+ a_7^q{\cal{T}}^{\mu\nu} 
  \bar{q}\sigma_{\mu\nu} q  
\right],
& &
\end{eqnarray}
%%%%%%%%%%%%%%%%%%%%%%%%%%%%%%%%%%%%%%%%%%%%%%%%%%%%%%%%%%
where the ${\cal{S},\cal{P}},
  {\cal{V}}^\mu,{\cal{A}}^\mu$ and ${\cal{T}}^{\mu\nu}$ are
  respectively linear combinations of scalar, 
  pseudoscalar, vector, 
  axial-vector, and tensor operators constructed from
  ${\cal{X}}$ and arbitrary 
  constants.  The constants $a_i^q$ are parametrized in terms of the
  Fermi constant
  $\gf=\sqrt{2}g^2/(8\mw^2)$  (where $\mw$ is the
  $W$-boson mass) and depend on the particular particle-physics
  model (such as the
  MSSM) used to generate the ${\cal{X}}$-quark Lagrangian.  The inclusion of
  $n$-body quark operators with $n>1$ would 
  induce further suppression in a heavy scale such as $\mw$ and are excluded.
In the following, we will neglect the $a_2^q$ term which is generally
  suppressed by an inverse power of the WIMP mass.  Also,
only SU(2) isospin symmetry will be considered since kaon-exchange
currents are suppressed in elastic WIMP-nucleus scattering.  We will
  therefore truncate
the sum  in \Eq{quarklag} to up and down quarks when considering
  vertices that contribute to exchange currents.

The Lagrangian of \Eq{quarklag} generates various ${\cal{X}}$-hadron
interactions vertices.  For the particular problem of ${\cal{X}}$-nucleus
scattering at NLO, we will focus on the following vertices:  
$\pi^2{\cal{X}}^2$, $\pi N^2 {\cal{X}}^2$
(which contribute to the two-nucleon current at NLO) and
$N^2{\cal{X}}^2$ (the one-nucleon contribution to the 
scattering amplitude).  
The corresponding ${\cal{X}}$-hadron Lagrangian will have the general form
%%%%%%%%%%%%%%%%%%%%%%%%%%%%%%%%%%%%%%%%%%%%%%%%%%%%%%%%%%
\begin{eqnarray}\label{general}
{\cal{L}}_{\cal{X}} = {\cal{L}}^{\pi\pi}+{\cal{L}}^{\pi N^2}+
{\cal{L}}^{N^2};
\end{eqnarray}
%%%%%%%%%%%%%%%%%%%%%%%%%%%%%%%%%%%%%%%%%%%%%%%%%%%%%%%%%%
the last of these is that which has been considered in prior work,
while the first two are new.  The three parts of ${\cal{L}}_{\cal{X}}$
in \Eq{general} are characterized by parameters $b_i,c_j,d_k$,
respectively, which in turn depend on $a_i^q$ of \Eq{quarklag}.
The precise shape of this general Lagrangian will therefore depend on
the ${\cal{X}}$-quark Lagrangian and we now proceed to construct each
term in \Eq{general} in turn.
\\

{\bf $\pi^2{\cal{X}}^2$ vertex. }
Up to NLO in powers of the pion momentum, the
$\pi^2{\cal{X}}^2$ Lagrangian will look like
%%%%%%%%%%%%%%%%%%%%%%%%%%%%%%%%%%%%%%%%%%%%%%%%%%%%%%%%%%
\begin{eqnarray}\label{pipixx}
{\cal{L}}^{\pi\pi} = b_{\text{s}} \vec{\boldmath\pi}\cdot\vec{\boldmath \pi}
{\cal{S}}
+ b_{\text{v}} i(\vec{\boldmath\pi}\times\partial^\mu\vec{\boldmath
  \pi})_3
{\cal{V}}_\mu .
\end{eqnarray}
%%%%%%%%%%%%%%%%%%%%%%%%%%%%%%%%%%%%%%%%%%%%%%%%%%%%%%%%%%
In \Eq{pipixx}, only the terms with two WIMPs will be kept in the
expansion of ${\cal{S},\cal{P}},{\cal{V}}^\mu$, and ${\cal{A}}^\mu$.
To derive an expression for $b_{\text{s}}$, we use the  matrix element,
%%%%%%%%%%%%%%%%%%%%%%%%%%%%%%%%%%%%%%%%%%%%%%%%%%%%%%%%%%
\begin{eqnarray}\label{matrixb0}
b_{\text{s}} \langle \pi^a |
\vec{\boldmath\pi}\cdot\vec{\boldmath \pi}
| \pi^a \rangle =
\gf \langle \pi^a |
\sum\limits_{q=u,d}
a_1^q \bar{q}q
| \pi^a \rangle ,
\end{eqnarray}
%%%%%%%%%%%%%%%%%%%%%%%%%%%%%%%%%%%%%%%%%%%%%%%%%%%%%%%%%%
from which we obtain using soft-pion techniques~\cite{Donoghue:dd},
%%%%%%%%%%%%%%%%%%%%%%%%%%%%%%%%%%%%%%%%%%%%%%%%%%%%%%%%%%
\begin{eqnarray}\label{b1}
b_{\text{s}}  =
 \frac{\gf m_\pi^2}{2(m_u+m_d)}(a_1^u+a_1^d) ,
\end{eqnarray}
%%%%%%%%%%%%%%%%%%%%%%%%%%%%%%%%%%%%%%%%%%%%%%%%%%%%%%%%%%
where the $m_i$ are current quark masses. The sum of the quark masses
is given in Ref.~\cite{Hagiwara:fs} as
$5\text{~MeV}<m_u+m_d<11 \text{~MeV}$ and we will use the average value
of $8\pm 3$~MeV.

For the NLO term of \Eq{pipixx}, we use the conservation of the vector
current (CVC) to write
%%%%%%%%%%%%%%%%%%%%%%%%%%%%%%%%%%%%%%%%%%%%%%%%%%%%%%%%%%
\begin{eqnarray}\label{b2}
b_v  = \gf(a_3^u - a_3^d) .
\end{eqnarray}
%%%%%%%%%%%%%%%%%%%%%%%%%%%%%%%%%%%%%%%%%%%%%%%%%%%%%%%%%%
Note that in the MSSM there is no contribution from $a_3^q$.
\\

{\bf $\pi N^2{\cal{X}}^2$ vertex.}
The NLO two-nucleon currents will only receive
contributions from the $a_{4,5}^q$ terms in \Eq{quarklag} (the other
possible terms contribute at NNLO) yielding 
%%%%%%%%%%%%%%%%%%%%%%%%%%%%%%%%%%%%%%%%%%%%%%%%%%%%%%%%%%
\begin{eqnarray}\label{pinnxx}
{\cal{L}}^{\pi N^2} = 
c_{\text{v}}\!\bar{N}\gamma^\mu i(\!\vec{\tau}\times\vec{\pi}\!)_3 N
{\cal{V}}_\mu
+ 
c_{\text{a}}\!\bar{N}\gamma^\mu\gamma^5 i(\!\vec{\tau}\times\vec{\pi}\!)_3 N
{\cal{A}}_\mu ,
& &
\end{eqnarray}
%%%%%%%%%%%%%%%%%%%%%%%%%%%%%%%%%%%%%%%%%%%%%%%%%%%%%%%%%%
where $N$ is the nucleon isospinor.  To extract the hadronic scales
that appear as one  
matches the quark-${\cal{X}}$ Lagrangian to the hadron-${\cal{X}}$
Lagrangian we can use dimensional analysis and the
scaling rule~\cite{Manohar:1983md,Prezeau:2003xn}  
%%%%%%%%%%%%%%%%%%%%%%%%%%%%%%%%%%%%%%%%%%%%%%%%%%%%%%%%%%%%%%%%%
\begin{equation}\label{georgi}
\left({ \frac{\bar{N}N}{\lh f_\pi^2} }\right)^k
\left({\frac{\partial^\mu}{\Lambda_{\text{H}}}}\right)^l
\left({\frac{\pi}{f_\pi}}\right)^m
\frac{\gf}{\lh}
\ \ \times (\Lambda_{\text{H}} f_\pi)^2,
\end{equation}
%%%%%%%%%%%%%%%%%%%%%%%%%%%%%%%%%%%%%%%%%%%%%%%%%%%%%%%%%%%%%%%%%
where $f_\pi\cong 92.4$~MeV is the pion decay constant and $(k,l,m)$
refer to the form of the hadronic part.  The
$c_i$'s of \Eq{pinnxx} which have one nucleon current and one pion
[hence, $(k,l,m)=(1,0,1)$] can be rewritten
%%%%%%%%%%%%%%%%%%%%%%%%%%%%%%%%%%%%%%%%%%%%%%%%%%%%%%%%%%
\begin{eqnarray}\label{cis}
 c_{\text{v,a}} = \frac{\gf}{f_\pi} (a_{4,5}^u-
 a_{4,5}^d)\delta_{\text{v,a}}~, 
\end{eqnarray}
%%%%%%%%%%%%%%%%%%%%%%%%%%%%%%%%%%%%%%%%%%%%%%%%%%%%%%%%%%
where $\delta_{\text{v}}$ is ${\cal{O}}(1)$ and from CVC we have
$\delta_{\text{a}}= -g_{\text{A}}/2$ with $g_{\text{A}}=1.27$, the
usual axial pion-nucleon coupling.  Note that in the MSSM,
$c_{\text{v}}=c_{\text{a}}=0$.
\\

{\bf $N^2{\cal{X}}^2$ vertex.}
The one-nucleon contribution to the scattering amplitude is
traditionally the only term considered and 
the full $N^2{\cal{X}}^2$ Lagrangian is given in
Ref.~\cite{kurylov03}.  In contrast to the two-nucleon case, the sum
in \Eq{quarklag} is now over all quark flavors.  At LO and
ignoring the pion-pole term for simplicity, we have
%%%%%%%%%%%%%%%%%%%%%%%%%%%%%%%%%%%%%%%%%%%%%%%%%%%%%%%%%%
\begin{eqnarray}\label{nrnnxx}
{\cal{L}}^{ N^2} \cong
{\cal{S}} \bar{N}(d_{\text{s}}^0 +d_{\text{s}}^1\tau^3)  N  +  {\cal{V}}_\mu 
 \bar{N} \gamma^\mu (d_{\text{v}}^0 +d_{\text{v}}^1\tau^3) N
~~~~~~~& &
\nonumber
\\
+\!  {\cal{A}}_\mu \!
 \bar{N}\! \gamma^\mu \gamma^5\!(\! d_{\text{a}}^0
 +d_{\text{a}}^1\tau^3\! )\! N
\!+\!  {\cal{T}}_{\mu\nu}\!
 \bar{N}\! \sigma^{\mu\nu}\! (\! d_{\text{t}}^0\!
 +\!d_{\text{t}}^1\tau^3\! )\! N,
~~& &
\end{eqnarray}
%%%%%%%%%%%%%%%%%%%%%%%%%%%%%%%%%%%%%%%%%%%%%%%%%%%%%%%%%%
where [with $(k,l,m)=(1,0,0)$]
%%%%%%%%%%%%%%%%%%%%%%%%%%%%%%%%%%%%%%%%%%%%%%%%%%%%%%%%%%
\begin{eqnarray}
d_{i}^0\!  =\!  \frac{\gf}{2}\! \left[\!\left(\! a_{j}^u \! +\!
a_{j}^d\!\right) \!\epsilon_{i}^0\! +\! 2\!\!
\sum\limits_{s,b,c,t}\! a_{j}^q\! \epsilon_{i}^q
\!\right]\! , ~
d_{\text{i}}^1\! =\! \frac{\gf}{2}\! \left(\!a_{j}^u\! -\! a_{j}^d\!
\right) \! \epsilon_{i}^1, ~
\end{eqnarray}
%%%%%%%%%%%%%%%%%%%%%%%%%%%%%%%%%%%%%%%%%%%%%%%%%%%%%%%%%%
and $j=\{1,3,6,7\}$ for $i=$ \{s,v,a,t\} respectively.  The
$\epsilon$'s are  ${\cal{O}}(1)$ except for $\epsilon_{\text{s}}^1=
\langle N| \bar{u}u-\bar{d}d |N\rangle$---where $N$ can be either a
neutron or a proton---which is small
numerically and will henceforth be neglected.  For the scalar terms,
we have in particular~\cite{Schweitzer:2003sb,Pavan:2001wz}
%%%%%%%%%%%%%%%%%%%%%%%%%%%%%%%%%%%%%%%%%%%%%%%%%%%%%%%%%%
\begin{eqnarray}\label{epsilons}
\epsilon_{\text{s}}^0 &=&
\langle N| \bar{u}u+\bar{d}d |N\rangle \cong 16 \pm 8 ,
\\
\epsilon_{\text{s}}^s &=& \langle N| \bar{s}s |N\rangle \approx (0.04-0.2)
\epsilon_{\text{s}}^0 ,
\\
\epsilon_{\text{s}}^{Q} &=& \langle N|\! \bar{Q}Q\! |N\rangle
= \frac{2}{27}\! \frac{M}{M_Q}\!
\left(\! 1 -\! \frac{m_u+m_d}{2M}\epsilon_{\text{s}}^0\! -\! 
\frac{m_s}{M}\epsilon_{\text{s}}^s
\!\right) ,
~~~~~
\end{eqnarray}
%%%%%%%%%%%%%%%%%%%%%%%%%%%%%%%%%%%%%%%%%%%%%%%%%%%%%%%%%%
where $Q=c,b,t$.  Although Ref.~\cite{Pavan:2001wz} suggests the larger
value for $\epsilon_{\text{s}}^s \sim 0.2 \epsilon_{\text{s}}^0 $, the
error is very substantial and we 
will use the more central value $\epsilon_{\text{s}}^s \sim 0.1
\epsilon_{\text{s}}^0$.
\\

{\bf MSSM.}  In the MSSM, the WIMP is the Majorana spin-1/2 neutralino
which couples to 
quarks according to the low-energy Lagrangian of \Eq{quarklag} with
$a_3^q=a_4^q=a_5^q=a_7^q=0$~\cite{vogel92}.   Since we are interested in
comparing the  
one- and two-nucleon scalar-scalar interactions of \Eq{quarklag}
we will not consider the axial-axial $a_6^q$ term and will only focus
on $a_1^q$ which in the MSSM is given by 
%%%%%%%%%%%%%%%%%%%%%%%%%%%%%%%%%%%%%%%%%%%%%%%%%%%%%%%%%%
\begin{eqnarray}\label{mssmaoneq}
a_1^q = 2\sqrt{2}\frac{m_q}{\mw}S_q,
\end{eqnarray}
%%%%%%%%%%%%%%%%%%%%%%%%%%%%%%%%%%%%%%%%%%%%%%%%%%%%%%%%%%
where $m_q$ is the
quark mass and the $S_q$'s are dimensionless and 
are usually larger than 10 in the MSSM~\cite{vogel92}.
The LO SI $\chi^2$-hadron Lagrangian generated by the scalar currents
is simply 
%%%%%%%%%%%%%%%%%%%%%%%%%%%%%%%%%%%%%%%%%%%%%%%%%%%%%%%%%%
\begin{eqnarray}\label{hadxx}
{\cal{L}}^{\chi\text{had}}_0 =
\left( d_{\text{s}}^0\bar{N}  N + b_{\text{s}} \vec{\pi}\cdot \vec{\pi}
\right)\bar{\chi}\chi  .
\end{eqnarray}
%%%%%%%%%%%%%%%%%%%%%%%%%%%%%%%%%%%%%%%%%%%%%%%%%%%%%%%%%%
Therefore, in the MSSM, an important
modification of the SI  $\chi^2$-hadron Lagrangian is the new
$\pi^2\chi^2$ term at LO.  This term,
through the two-nucleon diagram of \Fig{twobody}(a), can give a
contribution comparable in size to the one-nucleon scattering
amplitude.

The calculation of the diagram in \Fig{twobody}(a) is
straightforward~\cite{Prezeau:2003xn}:  
we use the Feynman rules to evaluate the amplitude for two particular
nucleons inside the nucleus and we take the
non-relativistic limit; we then Fourier transform the result and
obtain the operator
%%%%%%%%%%%%%%%%%%%%%%%%%%%%%%%%%%%%%%%%%%%%%%%%%%%%%%%
\begin{eqnarray}\label{approxa}
M_0^{\pi\pi} &\simeq&
i\frac{S_\pi m_\pi}{3\sqrt{2}M^2}
\frac{g_{\pi NN}^2}{4\pi}
\gf
~\bar{\chi}\chi
{\cal{O}}_0^{\pi\pi}(\vec{x}_1,\ldots,\vec{x}_4),
\end{eqnarray}
%%%%%%%%%%%%%%%%%%%%%%%%%%%%%%%%%%%%%%%%%%%%%%%%%%%%%%
where $M$ is the nucleon mass, $g_{\pi NN}^2/(4\pi)\cong 13.7$ is the
strong pion-nucleon coupling constant and the nuclear operator is
given by 
%%%%%%%%%%%%%%%%%%%%%%%%%%%%%%%%%%%%%%%%%%%%%%%%%%%%%%%
\begin{eqnarray}\label{lonuclop}
& &{\cal{O}}_0^{\pi\pi}(\vec{x}_1,\ldots,\vec{x}_4)
=-\delta(\vec{x}_1-\vec{x}_3) \delta(\vec{x}_2-\vec{x}_4)
\nonumber
\\
& & \times(\xi_{3,\alpha}^\dagger  \xi_{1,\beta})
(\xi_{4,\phi}^\dagger\xi_{2,\delta})
(\eta_{3,\sigma}^\dagger  \eta_{1,\epsilon})
(\eta_{4,\gamma}^\dagger\eta_{2,\zeta})
\nonumber
\\
& &\times
\frac{1}{x}
[\text{F}_1(x)\vec{\sigma}_{\alpha\beta}\cdot\vec{\sigma}_{\phi\delta}
+\text{F}_2(x) T_{\alpha\phi,\beta\delta}]
\vec{\tau}_{\sigma\epsilon}\cdot\vec{\tau}_{\gamma\zeta},
\end{eqnarray}
%%%%%%%%%%%%%%%%%%%%%%%%%%%%%%%%%%%%%%%%%%%%%%%%%%%%%%
where the $\xi$'s and $\eta$'s are spinors and isospinors
respectively, $~T_{\alpha\phi,\beta\delta}\equiv
3\vec{\sigma}_{\alpha\beta}\cdot\hat{\rho}
\vec{\sigma}_{\phi\delta}\cdot\hat{\rho}
- \vec{\sigma}_{\alpha\beta}\cdot\vec{\sigma}_{\phi\delta}~$ and
%%%%%%%%%%%%%%%%%%%%%%%%%%%%%%%%%%%%%%%%%%%%%%%%%%%%%%
\begin{eqnarray}
\text{F}_1(x)&=&(x-2)\text{e}^{-x},~~~
\text{F}_2(x)=(x+1)\text{e}^{-x},
\\
S_\pi &=& \frac{m_\pi^2}{(m_u+m_d)}\left(S_u\frac{m_u}{\mw}
+S_d\frac{m_d}{\mw}\right),
\end{eqnarray}
%%%%%%%%%%%%%%%%%%%%%%%%%%%%%%%%%%%%%%%%%%%%%%%%%%%%%%
where $x = m_\pi \rho$ is proportional to the distance between the two
nucleons with $\vec{\rho}=\vec{x}_1-\vec{x}_2$, $\rho=| 
\vec{\rho}|$ and ${\hat\rho}={\vec\rho}/\rho$.  In
\Eq{lonuclop}, 
$\vec{\sigma}$ and $\vec{\tau}$ are spin and isospin Pauli matrices
respectively.

The contribution to the WIMP-nucleus scattering will therefore be
%%%%%%%%%%%%%%%%%%%%%%%%%%%%%%%%%%%%%%%%%%%%%%%%%%%%%%
\begin{eqnarray}
{\cal{A}}^{\pi\pi}&=&i\frac{S_\pi m_\pi}{3\sqrt{2}M^2}
\frac{g_{\pi NN}^2}{4\pi}
\gf{\cal{N}}_0^{\pi\pi},
\\
{\cal{N}}_0^{\pi\pi}&=&\langle Z,A|\frac{1}{2}\sum\limits_{i\neq j}
{\cal{O}}_0^{\pi\pi}|Z,A\rangle,
\end{eqnarray}
%%%%%%%%%%%%%%%%%%%%%%%%%%%%%%%%%%%%%%%%%%%%%%%%%%%%%%
where the sum is over all nucleon pairs inside the nucleus, $A$ is
the number of nucleons, and $Z$ is the charge of the nucleus.

On the other hand, the one-nucleon contribution will simply be
%%%%%%%%%%%%%%%%%%%%%%%%%%%%%%%%%%%%%%%%%%%%%%%%%%%%%%
\begin{eqnarray}
{\cal{A}}^{NN}= 2i d_{\text{s}}^0 A .
\end{eqnarray}
%%%%%%%%%%%%%%%%%%%%%%%%%%%%%%%%%%%%%%%%%%%%%%%%%%%%%%
Taking the ratio of the
two-nucleon to one-nucleon contribution we obtain 
%%%%%%%%%%%%%%%%%%%%%%%%%%%%%%%%%%%%%%%%%%%%%%%%%%%%%%
\begin{eqnarray}\label{ratioofampls}
\frac{{\cal{A}}^{\pi\pi}}{{\cal{A}}^{NN}}\cong
(0.21 \pm 0.08) r \frac{{\cal{N}}_0^{\pi\pi}}{A},~
r \equiv \frac{S_u m_u+S_d m_d}{S^{NN}},
~~~~~~& &
\\
\label{mssmpara}
S^{NN}\equiv \frac{1}{2}(S_u m_u + S_d m_d) \epsilon_{\text{s}}^0
+\sum\limits_{s,c,b,t} S_q m_q \epsilon_{\text{s}}^q .
~~~~~~& &
\end{eqnarray}
%%%%%%%%%%%%%%%%%%%%%%%%%%%%%%%%%%%%%%%%%%%%%%%%%%%%%%
Note that we did not include contributions from heavy quarks 
with extra, inverse powers of the squark masses for
simplicity~\cite{kam-rev}, and that the error in the overall factor of
0.21 comes from the uncertainty of the current-quark masses.  Looking
at the special case where all the $S_q $ are equal and cancel in the 
ratio $r$ with $m_s=115\pm 55$~MeV, we find $0.002<(0.21\pm
0.08)r < 0.006 $.  Since there exists regions of the MSSM parameter
space where the $S_q$'s have opposite signs, the ratio $r$ can be
greater than one if cancellations occur.  For example, with
$S_u=S_d=S_s=-S_c=-S_b=-S_t$, $m_s\epsilon_{\text{s}}^s=115$~MeV, and
$(m_u+m_d)/2\epsilon_{\text{s}}^0=60$~MeV, $r\cong 1.5$.
Alternatively, if all the $S_q$'s have the same sign, a cancellation
could still occur with the extra heavy quark
terms not included in \Eq{mssmpara} if they are of similar
magnitudes but opposite in sign.

We estimated ${\cal{N}}_0^{\pi\pi}$ in the independent particle
approximation assuming closed shells and $N=Z$ where $N$ is the number
of neutrons.  We performed the calculation in two different mean-field
potentials: a square well and a harmonic-oscillator potential.  To
account for the hard core, we used Jastrow functions \{defined
  $[1-\text{exp}(-a\rho^2)(1-b\rho^2)]$ with $a=1.1~\text{fm}^{-2}$
and $b=0.68~\text{fm}^{-2}$\}  as well as a 
simple integral lower cut-off at 0.5~fm.  We found that the matrix
element is approximately linear in $A$ for nucleons in a harmonic
potential and exactly linear in $A$ for nucleons trapped in a square
well.  We also found a minimum value of 
${\cal{N}}_0^{\pi\pi}\cong A$ in the 
harmonic oscillator potential with Jastrow functions and a maximum
value of ${\cal{N}}_0^{\pi\pi}\cong 2A$ for the square-well potential
with the lower cut-off at 0.5~fm.  We estimate that
${\cal{N}}_0^{\pi\pi}/A\approx 1.5\pm 0.5$ although the error can be
significantly larger due to the fact that
the nuclei used in direct dark-matter searches are not closed-shell
nuclei and that they have a neutron excess.  We also do not take into
account spin-orbit coupling.  Furthermore, ${\cal{N}}_0^{\pi\pi}/A$
could vary substantially from one nucleus to the next.   Keeping these
issues in mind, we conclude that ${\cal{N}}_0^{\pi\pi}$ and $A$ are of
the same order and that the ratio in \Eq{ratioofampls} can be of the order
of one or larger depending on the values of the MSSM parameters.

In supersymmetric models, the one-nucleon current generically
produces roughly equal SI couplings to the proton and neutron
\cite{kurylov03}, which results in a SI amplitude that is 
proportional to the atomic number of the nucleus. Inclusion of the
two-nucleon contributions could change this 
picture since such contributions might cancel against the one-nucleon 
contributions. If the ratio of the two-nucleon
matrix element to the atomic number varies from one nucleus to the next
so will the degree of the cancellation. Thus, when the
two-nucleon contribution is taken into account, a dark-matter
candidate that appears in DAMA but not in other searches \cite
{dama-naI} is
conceivable for a WIMP with SI interactions even within the
framework of the MSSM (Ref. \cite{kam-vogel} provides arguments
against a WIMP with SD interactions as an explanation for the
DAMA signal).

With $1 < {\cal{N}}_0^{\pi\pi}/A < 2$ and $r\approx 1$, the
exchange-current contribution is in the range of 10-60\%.
This is atypical since such currents usually contribute about 5-10\%
to the amplitude as in $p+n\to d+\gamma$~\cite{Riska:1989bh} and
$\nu_l+  d\to \nu+ p+n $~\cite{Butler:1999sv}.
Part of the reason the exchange current contribution is normally
small at low-energies stems from the fact that the
``probe''-$\pi^2$ vertex (where the probe can be a photon or a neutrino for
example) is typically suppressed by at least a factor of $q/M\sim
0.1$.   In WIMP-nucleus scattering, the situation is different since the
$\pi^2{\cal{X}}^2$ vertex 
need not be suppressed by this factor as we showed
in \Eq{hadxx}. 
Thus, instead of a 5-10\% effect on the WIMP-nucleus scattering
amplitude, the one-nucleon and two-nucleon
contributions could be comparable.

In conclusion, we have shown that the two-nucleon 
and one-nucleon currents can give comparable contributions to the
WIMP-nucleus scattering and point out that this result can help
resolve the conflict between the various direct dark-matter search
experiments.

This work was supported in part by NASA NAG5-9821 and DoE
DE-FG03-92-ER40701 and DE-FG03-02ER41215.

%%%%%%%%%%%%%%%%%%%%%%%%%%%%%%%%%%%%%%%%%%%%%%%%%%%%%%%%%%%%%%%%%%%%

\end{document}